\documentclass[reprint,
 amsmath,amssymb,
aps,
pra,
]{revtex4-2}
\usepackage{mathrsfs}
\usepackage{caption}
\usepackage{graphicx}% Include figure files
\usepackage{dcolumn}% Align table columns on decimal point
\usepackage{bm}% bold math
\usepackage{float}
\usepackage{subfig}
\usepackage{color}
\usepackage{hyperref}% add hypertext capabilities
\begin{document}

\preprint{APS/123-QED}

\title{Nonclassical correlations and quadrature squeezing of photons in anisotropic quantum Rabi-Stark model}



	%\author{Jiong Li$^{1}$}
	%\author{Qing-Hu Chen$^{1,2,}$}
	%\email{qhchen@zju.edu.cn}
	
	%\affiliation{$^{1}$Zhejiang Key Laboratory of Micro-Nano Quantum Chips and Quantum Control, School of Physics, Zhejiang University, Hangzhou 310027, China\\
	%	$^{2}$Collaborative Innovation Center of Advanced Microstructures, Nanjing University, Nanjing 210093, China}

\author{Yong-Xin Zhang$^{1}$}
\author{Chen Wang$^{2,}$}
\email{wangchen@zjnu.cn}
\author{Qing-Hu Chen$^{1,3,}$}
\email{qhchen@zju.edu.cn}
\affiliation{$^{1}$Zhejiang Key Laboratory of Micro-Nano Quantum Chips and Quantum Control, School of Physics, Zhejiang University, Hangzhou 310027, China\\
                $^{2}$Department of Physics, Zhejiang Normal University, Jinhua 321004, China\\
		$^{3}$Collaborative Innovation Center of Advanced Microstructures, Nanjing University, Nanjing 210093, China}

\date{\today}

\begin{abstract}
This study investigates nonclassical effects of photons in the anisotropic quantum Rabi-Stark model by using a quantum dressed master equation. We analyze second- and higher-order correlation functions, and demonstrate that the nonlinear Stark coupling significantly modulates photon statistics, inducing distinct and tunable photon antibunching and bunching effects. Successive transition signatures of correlation functions provide a potential experimental probe for predicting quantum phase transitions. We further reveal that Stark coupling provides direct control over photon squeezing, achieving significant enhancement and suppression. These findings not only uncover rich nonclassical phenomena in the anisotropic quantum Rabi-Stark model, but also establish the nonlinear Stark coupling as a crucial new dimension for quantum detection. It may open a new avenue for precise manipulation of strongly coupled light-matter systems, with potential applications in quantum information processing and quantum-enhanced technologies.
\end{abstract}

%\keywords{}

\maketitle

\section{Introduction}\label{sec:intro}
 
Nonclassical light, a key branch of modern quantum optics, demonstrates quantum phenomena that violate classical wave theory, beginning with the landmark Brown and Twiss experiment of 1956 \cite{BROWN1956N_177_27-29}. This breakthrough birthed Glauber's quantum coherence theory~\cite{Glauber1963PR_130_2529-2539}, laying the foundation for characterizing nonclassical light. Subsequent discoveries of classically inhibited effects not only confirmed the quantum nature of light but also paved the way for quantum technologies. Artificially engineered nonclassical states are advancing quantum computing~\cite{Kok2007RMP_79_135-174,Raussendorf2007PRL_98_190504,Aaronson2011PFAASTC__333–342,Gambetta2017nQI_3_2,Liu2019NN_14_586-593,Wang2019NP_13_770-775}, secure communication \cite{Cleve1997PRA_56_1201-1204, Braunstein2005RMP_77_513-577, Kimble2008N_453_1023-1030, Scarani2009RMP_81_1301-1350, Wang2019PRA_99_042309}, and precision sensing \cite{Caves1980RMP_52_341-392b, Grangier1987PRL_59_2153-2156, Xiao1987PRL_59_278-281, Polzik1992PRL_68_3020-3023,  Aasi2013NP_7_613-619,Pedrozo-Penafiel2020N_588_414-418,LIGOScientificCollaborationandVirgoCollaboration2021PRX_11_021053,Grote2013PRL_110_181101}.
 
The advancement of photonic quantum technologies critically depends on the deterministic control of nonclassical effects, with photon blockade and squeezing serving as two key manifestations.  Specifically, photon blockade demonstrates temporal anti-correlations between photons, enabling efficient single-photon source~\cite{Kimble2008N_453_1023-1030,Scarani2009RMP_81_1301-1350} and quantum computing \cite{Kok2007RMP_79_135-174,Raussendorf2007PRL_98_190504,Aaronson2011PFAASTC__333–342,Gambetta2017nQI_3_2,Liu2019NN_14_586-593,Wang2019NP_13_770-775}. Meanwhile quadrature squeezing achieves noise reduction below the standard quantum limit, dramatically improving sensitivity in gravitational wave detection (e.g., LIGO) \cite{Aasi2013NP_7_613-619,LIGOScientificCollaborationandVirgoCollaboration2021PRX_11_021053,Grote2013PRL_110_181101} and quantum metrology \cite{Motes2015PRL_114_170802,Moller2017N_547_191-195,Colombo2022NP_18_925-930,Qin2023PRL_130_070801}. Current research explores diverse physical implementations, including optical parametric amplifiers \cite{Sarma2017PRA_96_053827,Shen2020PRA_101_013826,Wang2020OLO_45_2604-2607,Qin2021PRL_127_093602,Qin2022PRL_129_123602,Lou2023QST_8_045027,Singh2023QIP_22_198}, optomechanical resonators \cite{Liao2011PRA_83_033820,Rabl2011PRL_107_063601,Kronwald2013PRA_88_063833,Wollman2015S_349_952-955,Zhang2015PRA_91_063836,Nielsen2017PNAS_114_62-66,Wang2019PRA_99_043818,Schmidt2021QST_6_034005,Zheng2025QST_10_035005}, nonlinear optical cavities \cite{Leonski1994PRA_49_R20-R23,Nunnenkamp2010PRA_82_021806,Majumdar2013PRB_87_235319a,Xu2014PRA_90_043822,Lu2015PRA_91_013834,Sarma2018JPBAMOP_51_075505,Shi2019SR_9_8754} and cavity quantum electrodynamics systems \cite{Rebic2004PRA_69_035804,Birnbaum2005N_436_87-90,Dayan2008S_319_1062-1065,Faraon2008NP_4_859-863,Munoz2014NP_8_550-555,Tutunnikov2025QST_10_025002}. 
These quantum realizations lay the foundation for scalable implementation in quantum information technologies.
  
The quantum Rabi model (QRM)~\cite{Rabi1936PR_49_324-328,Scully1997__,Braak2016JPAMT_49_300301}, which describes the fundamental interaction between a two-level system and a single-mode bosonic field, exhibits distinctive nonclassical effects, including anomalous photon blockade \cite{Ridolfo2013PRL_110_163601,Chen2022JPBAMOP_55_115502} and quantum squeezing \cite{Genoni2015NJP_17_013034,Chen2022JPBAMOP_55_115502}. When the rotating-wave (RW) terms and counter-rotating-wave (CRW) counterparts have unequal weights, the QRM is extended to anisotropic QRM (AQRM). The AQRM exhibits first-order quantum phase transitions (QPTs) \cite{Xie2014PRX_4_021046,Liu2017PRL_119_, Chen2021PRA_103_043708} and additional photon antibunching at finitetemperatures~\cite{Ye2023PSMIA_609_128364,Ye2024OE_32_33483}, which are not present in the standard QRM \cite{Ridolfo2013PRL_110_163601,Hwang2015PRL_115_180404}. %Such anisotropic qubit-photon interactions provide a powerful route to manipulate photon blockade and other nonclassical effects \cite{Lu2022NJP_24_053029,Falch2025PRB_111_}. 
Notably, Cong \emph{et~al.}~\cite{congSelectiveInteractionsQuantum2020} proposed implementing AQRM in a trapped-ion system. They engineered Stark interactions using laser-driven transitions and achieved tunable anisotropic coupling by precisely controlling the red- and blue-sideband amplitudes. This breakthrough in quantum simulations has generated considerable theoretical interest. 

Recent studies have extensively explored the inclusion of the nonlinear Stark term in the QRM~\cite{Grimsmo2013PRA_87_033814,Grimsmo2014PRA_89_033802}. Due to Stark coupling, the extended quantum Rabi-Stark model (QRSM) exhibits singular behavior, including first-order QPTs and spectral collapse~\cite{Eckle2017JPAMT_50_294004,Xie2019JPAMT_52_245304,Chen2020PRA_102_}. Using von Neumann entropy calculations, Boutakka \textit{et al.} demonstrated that the nonlinear Stark term plays a critical role in modulating quantum entanglement \cite{Boutakka2025APB_131_113}. In addition to these effects, nonlinear Stark coupling has been shown to exhibit electromagnetically induced transparency, significantly improving system stability and efficiency \cite{Tang2019PRA_12_,Peng2022PRA_105_,Tang2022NJP_24_123021,Gou2024NJP_26_073046}. Considering the combination of AQRM and the nonlinear Stark term, Xie \textit{et al.} demonstrated that the anisotropic quantum Rabi-Stark model (AQRSM) exhibits both continuous and first-order QPTs across significantly  expanded parameter ranges ~\cite{xieFirstorderContinuousQuantum2020}, enhancing capabilities for quantum precision measurement. Furthermore, Xu~\textit{et~al.} showed that AQRSM-based quantum thermal machines achieve higher efficiency and power output than conventional harmonic oscillator systems~\cite{xuExploringRoleCriticality2024}. These results clearly establish the superior quantum control properties of AQRSM, opening new opportunities to study nonclassical effects, such as controlled photon blockade and quadrature squeezing. Building on these advancements, our research systematically investigates how Stark coupling influences the nonclassical behavior of photons within the AQRSM, elucidating the modulatory role of the nonlinear Stark coupling on these quantum effects.
    
To account for realistic environmental effects, our investigation employs the open quantum system framework \cite{Weiss2021__} and uses the dressed master equation (DME) approach~\cite{Beaudoin2011PRA_84_043832} to handle the strongly coupled AQRSM. At finite temperatures, we characterize steady-state nonclassical behaviors within tunable parameters, revealing Stark term's control over photon statistics~\cite{Ridolfo2012PRL_109_193602,Ridolfo2013PRL_110_163601}. By analyzing the second- and higher-order correlation functions, we uncover multiparameter control of photon bunching and antibunching transitions. The results establish a direct link to zero-temperature phase boundaries, providing an alternative approach for detecting QPTs. Additionally, quadrature squeezing \cite{Ma2011PR_509_89-165} is quantified across extended parameter regimes, demonstrating critical control of the nonlinear Stark coupling. These results provide a systematic framework for controlling photon nonclassicality through the nonlinear Stark coupling in strongly coupled light-matter open quantum systems.
    
The remainder of this paper is organized as follows. In Sec.~\ref{sec:Model and method}, we present a theoretical description of the AQRSM and DME approach in the context of open quantum systems. Sec.~\ref{sec:statistics} employs second- and higher-order correlation functions to systematically analyze the impact of the nonlinear Stark term on photon bunching and antibunching phenomena, establishing connections between these quantum statistical behaviors and QPT features. Sec.~\ref{sec:squeezing} examines the effect of nonlinear Stark coupling on quadrature squeezing through extensive parameter-space analyses, revealing the underlying physical mechanism for precise control. Finally, Sec.~\ref{sec:Conclusion} summarizes key findings and discusses promising directions for future research.

\section{Model and method}\label{sec:Model and method}
\subsection{Anisotropic quantum Rabi-Stark model}
The AQRSM describes a qubit anisotropically coupled to a single-mode cavity field. This system is characterized by following Hamiltonian
\begin{equation}
\begin{aligned}
H_{\textnormal{AQRSM}}=&(\frac{1}{2}\Delta +Ua^{\dag }a)\sigma _{z}+\omega _{0}a^{\dag
}a\\
&+g[(a\sigma _{+}+a^{\dag }\sigma _{-})+r(a\sigma _{-}+a^{\dag }\sigma
_{+})].
\end{aligned}
\label{H_AQRSM}
\end{equation}
where $\Delta $ denotes the qubit energy splitting, $U$ is the nonlinear Stark coupling strength, and $a^{\dag }(a)$ is the photonic creation (annihilation) operator for the cavity field with frequency $\omega _{0}$. $\sigma _{\pm }=\frac{1}{2}(\sigma _{x}\pm i\sigma _{y})$ are the qubit exciting and relaxing operators between the ground state $\left\vert 0\right\rangle $
and the excited state $\left\vert 1\right\rangle $, with the Pauli operators $\sigma _{\alpha=x,y,z}$. The parameter $g$ denotes the qubit-cavity coupling strength, while $r$ is the anisotropic parameter that controls the contribution of the CRW terms. When $U=0$, AQRSM reduces to AQRM~\cite{Xie2014PRX_4_021046,Liu2017PRL_119_, Chen2021PRA_103_043708}. If $r=1$, AQRSM reduces to the standard QRM~\cite{Rabi1936PR_49_324-328,Scully1997__,Braak2016JPAMT_49_300301}. In the weak-coupling limit, the CRW terms $(a\sigma_{-}+a^{\dag }\sigma _{+})$ are negligible, and AQRSM simplifies to the Jaynes-Cummings model (JCM)~\cite{Jaynes1963PI_51_89-109}, corresponding to $r=0$. With the experimental realization of strong coupling \cite{Clarke2008N_453_1031-1042}, ultra-strong coupling (USC) \cite{Forn-Diaz2010PRL_105_237001}, and deep-strong coupling (DSC) \cite{Casanova2010PRL_105_263603}, the CRW terms play an increasingly important role.

The system possesses $Z_{2}$ symmetry, as evidenced by the commutation relation $[\mathit{\Pi},H_{\textnormal{AQRSM}}]=0$. The conserved-parity operator $\mathit{\Pi}$ is defined as $\exp (i\pi N)$, where $N=a^{\dag }a+\frac{1}{2}(\sigma _{z}+1)$ is the total excitation number. The eigenvalues of $\mathit{\Pi}$ are $\pm 1$, depending on whether $N$ is odd or even. Although CRW terms break the conservation of the total excitation, the preserved parity symmetry not only facilitates theoretical analysis but also plays a key role in modulating nonclassical effects. All numerical results were obtained using a properly converged truncation of the photon Fock space at $N_{\textnormal{tr}}=200$. This parameter will not be restated in subsequent discussions.
\subsection{Quantum dressed master equation}
In realistic systems, quantum coherence is inevitably influenced by the environment \cite{Weiss2021__}. To model the dissipative effects in AQRSM, we couple both the qubit and the cavity field to separate thermal baths. The total Hamiltonian is expressed as
\begin{equation}
H_{\textnormal{total}}=H_{\textnormal{AQRSM}}+H_{\textnormal{B}}+V.  \label{H_total}
\end{equation}

The first term represents the system's Hamiltonian, as given in Eq.~\eqref{H_AQRSM}. The thermal baths are described by the Hamiltonian $H_{\textnormal{B}}=\sum_{u=\textnormal{q,c};k}\omega_{k}b_{u,k}^{\dag }b_{u,k}$, where $b_{u,k}^{\dag }$ ($b_{u,k}$) are the creation (annihilation) operators for bosonic modes with frequency $\omega _{k}$ in the $u$th thermal bath. The system-bath coupling is given by $V=V_{\textnormal{q}}+V_{\textnormal{c}}$, where $V_{\textnormal{q}}$ and $V_{\textnormal{c}}$ describe the interactions of the qubit and cavity field with their respective thermal baths. The specific forms of interactions are given by: 
\begin{subequations}
\begin{align}
V_{\textnormal{q}}=&\sum\limits_{k}\lambda _{\textnormal{q},k}(b_{\textnormal{q},k}+b_{\textnormal{q},k}^{\dag })\sigma
_{x},\\
V_{\textnormal{c}}=&\sum\limits_{k}\lambda _{\textnormal{c},k}(b_{\textnormal{c},k}+b_{\textnormal{c},k}^{\dag })(a+a^{\dag
}),  \label{Interaction}
\end{align}%
\end{subequations}
where $\lambda _{\textnormal{q}(\textnormal{c}),k}$ denotes the coupling strength between the qubit (cavity field) and the corresponding bosonic thermal bath. Generally, the coupling between each subsystem and its thermal bath is typically described by independent spectral functions, i.e., $J_{\textnormal{q}(\textnormal{c})}(\omega )=2\pi \sum_{k}|\lambda
_{\textnormal{q}(\textnormal{c}),k}|^{2}\delta (\omega -\omega _{k})$. This study adopts Ohmic-type spectral functions $J_{\textnormal{q}}(\omega )=\alpha _{\textnormal{q}}\frac{\omega }{\Delta }\textnormal{exp}(-\frac{\left\vert \omega\right\vert }{\omega _{c}})$ and $J_{\textnormal{c}}(\omega )=\alpha _{\textnormal{c}}\frac{\omega }{\omega _{0}}\textnormal{exp}(-\frac{\left\vert \omega
\right\vert }{\omega _{c}})$,
where $\alpha _{\textnormal{q}(\textnormal{c})}$ is the corresponding coupling strength and $\omega _{c}$ is the cutoff frequency of baths.

Under the assumption of weak system-bath coupling, the Born-Markov approximation is used to perturbatively handle $V_{\textnormal{c}}$ and $V_{\textnormal{q}}$, leading to the quantum DME \cite{Beaudoin2011PRA_84_043832}. This method can describe long-time dissipative dynamics, even in the USC and DSC regimes. In these regimes, the qubit-photon interaction is so strong that the system must be considered an indivisible whole, yet the quantum DME approach remains effective. Therefore, the system-bath interaction must be described under the eigenmodes $\left\vert \varphi _{n}\right\rangle $ of AQRSM, namely,
\begin{subequations}
\begin{align}
V_{\textnormal{q}}=&\sum_{k,m,n}\lambda _{\textnormal{q},k}(b_{\textnormal{q},k}+b_{\textnormal{q},k}^{\dag })P_{nm}^{\textnormal{q}}, \\
V_{\textnormal{c}}=&\sum_{k,m,n}\lambda _{\textnormal{c},k}(b_{\textnormal{c},k}+b_{\textnormal{c},k}^{\dag })P_{nm}^{\textnormal{c}},
\end{align}%
\end{subequations}
where
$P_{nm}^{\textnormal{q(\textnormal{c})}}$ is the projection operator defined as $P_{nm}^{\textnormal{q}}=\left\langle
\varphi _{n}\right\vert \sigma _{x}\left\vert \varphi _{m}\right\rangle
\left\vert \varphi _{n}\right\rangle \left\langle \varphi _{m}\right\vert $
and $P_{nm}^{\textnormal{c}}=\left\langle \varphi _{n}\right\vert (a^{\dag }+a)\left\vert
\varphi _{m}\right\rangle \left\vert \varphi _{n}\right\rangle \left\langle
\varphi _{m}\right\vert $. The quantum DME is thus given by %
\begin{equation}
\begin{aligned}
\frac{\partial \rho }{\partial t}=&-i[H_{\textnormal{AQRSM}},\rho]
+\sum\limits_{j,k>j}^{u=\textnormal{q,c}}\{\Gamma _{u}^{j,k}n_{u}(\Delta _{k,j})\mathcal{D}%
[\left\vert \varphi _{k}\right\rangle \left\langle \varphi _{j}\right\vert
,\rho ]\\
&+\Gamma _{u}^{j,k}[1+n_{u}(\Delta _{k,j})]%
\mathcal{D}[\left\vert \varphi _{j}\right\rangle \left\langle \varphi
_{k}\right\vert ,\rho]\} .
\end{aligned}
\label{DME}
\end{equation}%
Here, $\mathcal{D}[\mathcal{O},\rho ]=\frac{1}{2}(2\mathcal{O}\rho \mathcal{O}^{\dag
}-\mathcal{O}^{\dag }\mathcal{O}\rho-\rho \mathcal{O}^{\dag }\mathcal{O})$ represents the dissipator, and $\rho $ is the reduced density matrix of the system. The thermal occupation number is governed by the Bose-Einstein distribution $\ n_{u}(\Delta _{k,j})={1}/{[\textnormal{exp}(\Delta
_{k,j}/k_{\textnormal{B}}T_{u})-1]}$, where $k_{\textnormal{B}}$ is the Boltzmann constant and $T_{u}$ is the temperature of the $u$th thermal bath. $\Delta _{k,j}=E_{k}-E_{j}$ is the energy gap between the eigenstates $\left\vert
\varphi _{k}\right\rangle $ and $\left\vert \varphi _{j}\right\rangle $. The transition rates $\Gamma_{\textnormal{q}}^{k,j}$ and $\Gamma _{\textnormal{c}}^{k,j}$ between states $\left\vert \varphi
_{k}\right\rangle $ and $\left\vert \varphi _{j}\right\rangle $ are given by:
\begin{subequations}
\begin{align}
\Gamma _{\textnormal{q}}^{k,j} =&  \alpha _{\textnormal{q}}\frac{\Delta _{k,j}}{\Delta }\textnormal{exp}(-\frac{%
\left\vert \Delta _{k,j}\right\vert }{\omega _{c}})\left\vert \left\langle
\varphi _{j}\right\vert (\sigma _{-}+\sigma _{+})\left\vert \varphi
_{k}\right\rangle \right\vert ^{2},\\
\Gamma _{\textnormal{c}}^{k,j} =& \alpha _{\textnormal{c}}\frac{\Delta
_{k,j}}{\omega _{0}}\textnormal{exp}(-\frac{%
\left\vert \Delta _{k,j}\right\vert }{\omega _{c}})\left\vert \left\langle \varphi _{j}\right\vert (a+a^{\dagger
})\left\vert \varphi _{k}\right\rangle \right\vert ^{2}.  
\end{align}
\end{subequations}
As a result, parity conservation imposes selection rules on transitions, ensuring that the spin and photon ladder operators couple only states of opposite parity.

Using the quantum DME, we can determine the time evolution of each density matrix element. For the element $\rho_{m,n}$, its dynamics are governed by
\begin{eqnarray}
\begin{aligned}
\frac{\partial \rho _{m,n}}{\partial t} =&-i\Delta _{m,n}\rho
_{m,n}+\{\sum\limits_{k>m}^{u=\textnormal{q,c}}\Gamma _{u}^{m,k}[1+n_{u}(\Delta
_{k,m})]\\&+\sum\limits_{k<m}^{u=q,c}\Gamma _{u}^{k,m}n_{u}(\Delta
_{m,k})\}\delta _{m,n}\rho _{k,k} \\
&-\frac{1}{2}\{\sum\limits_{k<m}^{u=\textnormal{q,c}}\Gamma _{u}^{k,m}[1+n_{u}(\Delta
_{m,k})]\\
&+\sum\limits_{k<n}^{u=\textnormal{q,c}}\Gamma _{u}^{k,n}[1+n_{u}(\Delta
_{n,k})]\\
&+\sum\limits_{k>m}^{u=\textnormal{q,c}}\Gamma _{u}^{m,k}n_{u}(\Delta_{k,m})+\sum\limits_{k>n}^{u=\textnormal{q,c}}\Gamma _{u}^{n,k}n_{u}(\Delta
_{k,n})\}\rho _{m,n}.
\end{aligned}
\end{eqnarray}%
For $m\neq n$, it is general to find that $\rho _{m,n}=0$ is satisfied in the steady state. But for $m=n$, using the relations $1+n_{u}(\Delta
_{k,m})=-n_{u}(\Delta _{m,k})$ and $\Gamma _{u}^{m,k}=-\Gamma _{u}^{k,m}$.
The steady-state condition $\frac{\partial\rho ^{SS}}{\partial t}=0$ can be simplified as 
\begin{eqnarray}
\begin{aligned}
&\sum\limits_{k\neq m}^{u=\textnormal{q,c}}\Gamma _{u}^{k,m}n_{u}(\Delta _{m,k})\rho_{k,k}\\
&-\sum\limits_{k\neq m}^{u=\textnormal{q,c}}\Gamma _{u}^{k,m}[1+n_{u}(\Delta_{m,k})]\rho _{m,m}=0. \label{rou_f}
\end{aligned}
\end{eqnarray}%
When the bath temperatures are equal, $T_{\textnormal{q}}=T_{\textnormal{c}}=T$, the AQRSM reaches thermal equilibrium. In this state, the density matrix adopts the canonical ensemble form $\rho=\sum\limits_{n}P_n\left\vert \varphi
_{n}\right\rangle \left\langle \varphi _{n}\right\vert$, with probability $P_n=\rm{e}^{-\mathit{E_{n}}/\mathit{k}_{\textnormal{B}}\mathit{T}}/\rm{Z}$ and partition function $\rm{Z}=\rm{Tr}(\rm{e}^{-\mathit{H}_{\rm{AQRSM}}/{\mathit{k}_{\rm{B}}\mathit{T}}})$.
In the zero-temperature limit $T=0$, the steady-state solution dictated by Eq.~\eqref{DME} reduces to the system's ground state.

\section{Nonclassical correlations of photons}\label{sec:statistics}
Coherence has long been a fundamental concept in optics \cite{Glauber1963PR_130_2529-2539,Scully1997__}, with its definition evolving over time. The advent of quantum optics deepened our understanding of coherence, revealing that it encompasses not only classical interference, but also the quantum component in photon-involved processes \cite{BROWN1956N_177_27-29}. When the electromagnetic field is quantized, quantum coherence can be characterized by correlation functions of arbitrary order, which allows the characterization of nonclassical light fields. 
A key measurement is the two-photon correlation function~\cite{Scully1997__}.
When photons arrive at detectors at regular intervals rather than in clusters, as reflected by a second-order correlation function $0<G_{2}(0)<1$, photon antibunching, also known as single-photon blockade, occurs.

\subsection{Correlation functions in dressed-atom picture}
Experimentally, the coherence properties of quantum light fields are probed by measuring their quantum correlation functions \cite{Birnbaum2005N_436_87-90,Dayan2008S_319_1062-1065,Faraon2008NP_4_859-863}. The $n$th-order correlation function is typically defined as \cite{Glauber1963PR_130_2529-2539}
\begin{equation}
\begin{aligned}    
&G_{n}(\tau _{1},...,\tau _{n-1})\\&=\frac{\left\langle a^{\dag }(t)...a^{\dag }(t+\tau _{n-1})a(t+\tau _{n-1})...a(t)\right\rangle }{\left\langle (a^{\dag }a)(t)\right\rangle ...\left\langle (a^{\dag }a)(t+\tau _{n-1})\right\rangle },  \label{gn}
\end{aligned}
\end{equation}%
where $a(t)$ ($a^{\dag }(t)$) represents the field annihilation (creation) operator in the Heisenberg picture. In the dissipative AQRSM, the intracavity photon field couples to an external probe mode to measure emitted photons. When the qubit-cavity coupling is sufficiently weak, the emitted field can be approximated as proportional to the intracavity field, following standard input-output relations. Thus, the correlation properties of the emitted field in steady state can be reasonably described by the Eq.~\eqref{gn} function defined above. However, in the USC regime, this approximation of the emission field breaks down. This occurs because the approximation predicts a non-zero photon emission flux from the ground state of the coupled system, which contradicts experimental observations~\cite{Ridolfo2012PRL_109_193602}. Consequently, a more rigorous theoretical framework must be developed to accurately describe photon emission characteristics in the USC regime.

To analyze two-photon statistics in the USC regime,
Ridolfo \textit{et al.} \cite{Ridolfo2012PRL_109_193602,Ridolfo2013PRL_110_163601} developed a dressed-state framework with a modified
input-output relation $A_{\mathrm{out}}=A_{\mathrm{in}}-i\sqrt{\kappa }X^{+}$, where $%
A_{\mathrm{in(out)}}$ represents the input (output) radiation field operator, $\kappa 
$ is the photon leakage rate into external detection modes, and $X^{+}$ is the
dressed-state detection operator defined by $X^{+}=-i\sum_{j,k>j}\Delta
_{k,j}X_{j,k}\left\vert \varphi _{j}\right\rangle \left\langle \varphi
_{k}\right\vert $. Here, $X_{j,k}=\left\langle \varphi _{j}\right\vert
(a+a^{\dag })\left\vert \varphi _{k}\right\rangle$ are the matrix elements of the detection operator. For a vacuum input field, the output photon flux of the resonator is given by $I_{\mathrm{out}}=\kappa \left\langle X^{-}(t)X^{+}(t)\right\rangle $ with $X^{-}=(X^{+})^{\dag}$. This formulation resolves the unphysical prediction of the finite-photon current in the ground state, namely, $X^{+}\left\vert \varphi _{0}\right\rangle =0$. Crucially, the matrix elements $X_{j,k}$ and $X_{k,j}$ vanish when the eigenstates $\left\vert \varphi _{j}\right\rangle $ and $\left\vert \varphi
_{k}\right\rangle $ of the AQRSM have the same parity. This parity-matching condition serves as a selection rule for the quantum transitions. As a result, the $n$th-order photon correlation function must
be generalized as
\begin{equation}
\begin{aligned}
&G_{n}(\tau _{1},...,\tau _{n-1})\\
&=\frac{\left\langle X^{-}(t)...X^{-}(t+\tau _{n-1})X^{+}(t+\tau _{n-1})...X^{+}(t)\right\rangle }{\left\langle (X^{-}X^{+})(t)\right\rangle ...\left\langle(X^{-}X^{+})(t+\tau _{n-1})\right\rangle },  \label{Gn}
\end{aligned}
\end{equation}%
where the expectation value $\left\langle A(t)\right\rangle =Tr\{\rho
A(t)\}$ is calculated using the system's reduced density matrix $\rho $%
. Photon correlations typically decay to their uncorrelated value of 1 at infinite time delays, marking the equal-time correlation function  particularly significant for steady-state characterization. In this regime, the general correlation function $G_{n}(\tau _{1},...,\tau _{n-1})$ simplifies to its zero-time delay form as
\begin{equation}
G_{n}(0)=\frac{\left\langle X^{-n}X^{+n}\right\rangle _{ss}}{\left\langle
X^{-}X^{+}\right\rangle _{ss}^{n}},  \label{Gn(0)}
\end{equation}%
where $\left\langle ...\right\rangle _{ss}$ represents the expectation value in steady state.

Among the zero-time delay correlations, the two-photon correlation function $
G_{2}(0)$ serves as the most important indicator of nonclassical photon
statistics, expressed as
\begin{equation}
G_{2}(0)=\frac{\left\langle X^{-}X^{-}X^{+}X^{+}\right\rangle _{ss}}{%
\left\langle X^{-}X^{+}\right\rangle _{ss}^{2}}.  \label{G2(0)}
\end{equation}
This function reveals distinct quantum statistical behaviors: $G_{2}(0)>1$ indicates photon bunching, $G_{2}(0)<1$ signifies photon antibunching, and $G_{2}(0)=1$ corresponds to coherent states, reflecting their uncorrelated nature. 

The third-order correlation function $G_{3}(0)$ represents the first
higher-order coherence in quantum optics, characterizing the joint detection
probability of three photons, and is given by
\begin{equation}
G_{3}(0)=\frac{\left\langle X^{-}X^{-}X^{-}X^{+}X^{+}X^{+}\right\rangle _{ss}%
}{\left\langle X^{-}X^{+}\right\rangle _{ss}^{3}}.  \label{G3(0)}
\end{equation}
Compared to the second-order function $G_{2}(0)$, $G_{3}(0)$ is more sensitivity to photon statistics, allowing the detection of higher-order photon bunching or antibunching. Furthermore, it is directly related to multiphoton emission processes \cite{Munoz2014NP_8_550-555,Gou2024NJP_26_073046}, making it a powerful tool for investigating strongly coupled light-matter interactions.

\subsection{Tunable two-photon correlation via Stark coupling for Probing criticality}
In this subsection, we first explore the influence of the nonlinear Stark coupling $U$ on the steady state two-photon correlation function $G_{2}(0)$, as described by Eq.~\eqref{G2(0)}. We then investigate the intrinsic mechanism through which the Stark term influences the antibunching effect. Finally, we systematically examine the role of the second-order correlation function in predicting phase transitions influenced by the Stark term. 

\paragraph{Rich photon bunching and antibunching patterns}

\begin{figure}[!htbp]
    \centering
\includegraphics[width=8.6cm]{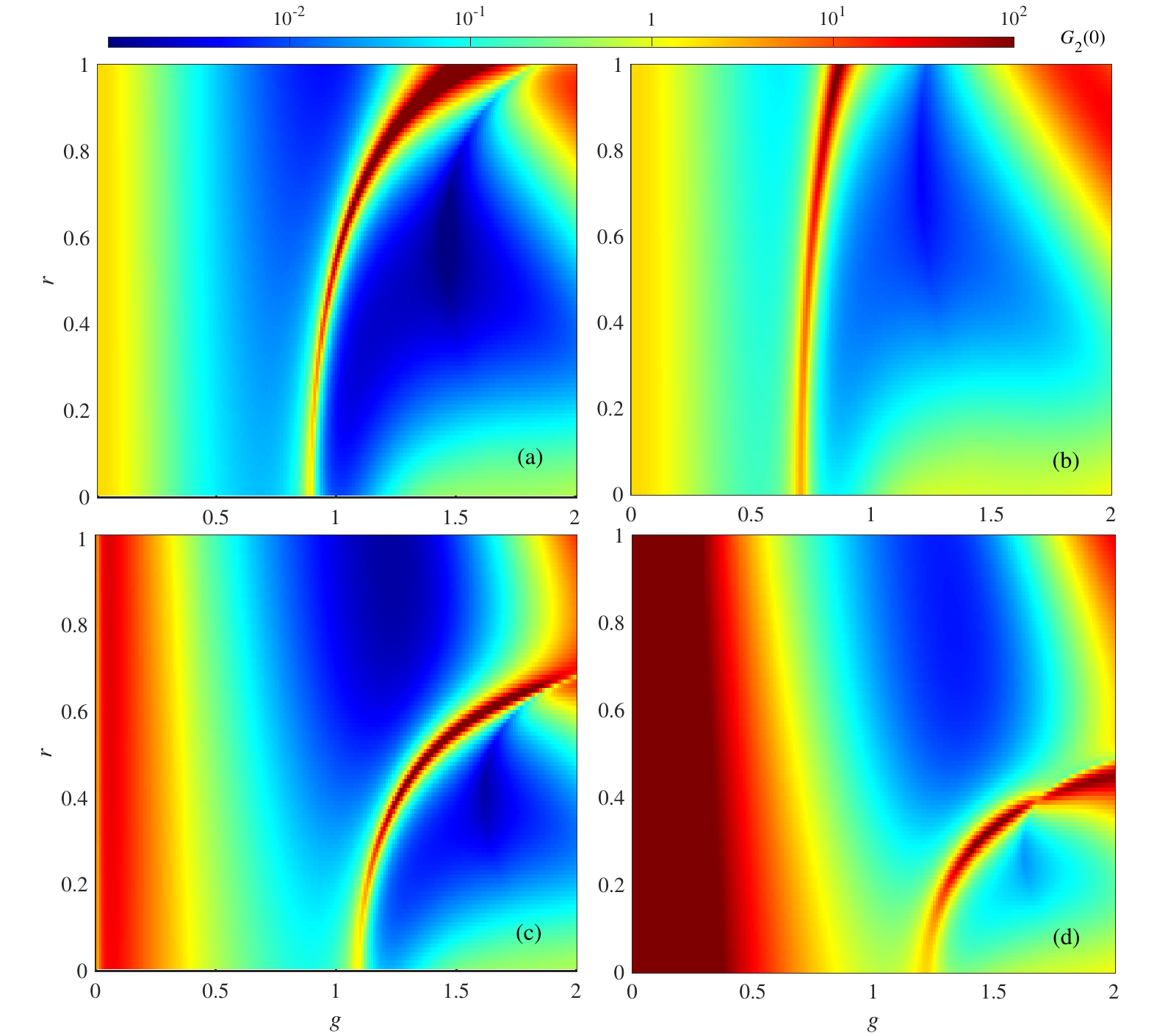}
    \vspace{-0.2cm}
    \caption{The zero-time delay two-photon correlation function,  $G_2(0)$, as a function of qubit-photon coupling strength $g$ and anisotropic parameter $r$, for distinct Stark coupling strengths: (a) $U=0.2$, (b) $U=0.5$, (c) $U=-0.2$, and $U=-0.5$. The other system parameters are  $\Delta=1$, $\omega_c=10\omega_0$, $\alpha_\mathrm{c}=\alpha_\mathrm{q}=10^{-3}$, and  $k_{\mathrm{B}}T_{\mathrm{c}}=k_{\mathrm{B}}T_{\mathrm{q}}=0.07\omega_{0}$.}
    \label{fig1}
\end{figure}

The second-order correlation function $G_{2}(0)$ in AQRSM exhibits distinct behaviors as the anisotropic parameter $r$ and the Stark coupling $U$ are tuned, as depicted in Fig. \ref{fig1}. The photon statistics transition from a double-switching pattern (antibunching $\to$ bunching $\to$ antibunching $\to$ bunching) to a single-switching pattern (antibunching $\to$ bunching), illustrating how both the anisotropy and the nonlinear Stark couplings modulate and enrich the photon statistics in dissipative qubit-photon interaction systems.  

When $U$ becomes positive, the nonlinear Stark coupling significantly broadens  the  parameter space for rich antibunching behavior, as shown in Figs.~\ref{fig1}(a)-(b), compared to the $U=0$ case~\cite{Ye2023PSMIA_609_128364,Ye2024OE_32_33483}. As $U$ increases, a secondary antibunching transition appears, even in regions where $r\rightarrow 1$, as shown in Fig.~\ref{fig1}(a), a transition not possible in the standard AQRM. Furthermore, in the regime dominated by CRW terms ($r>1$), the system still exhibits a double antibunching $\rightarrow$ bunching transition, as shown in Fig.~\ref{fig1}(b), which is strictly forbidden in the standard AQRM. This suggests  that the nonlinear Stark term diversifies the behavior of $G_{2}(0)$, enabling multiparameter control of quantum correlations and potentially enhancing the flexibility of quantum information processing \cite{Kok2007RMP_79_135-174,Raussendorf2007PRL_98_190504,Aaronson2011PFAASTC__333–342,Gambetta2017nQI_3_2,Liu2019NN_14_586-593,Wang2019NP_13_770-775,Cleve1997PRA_56_1201-1204, Braunstein2005RMP_77_513-577, Kimble2008N_453_1023-1030, Scarani2009RMP_81_1301-1350, Wang2019PRA_99_042309,Caves1980RMP_52_341-392b, Grangier1987PRL_59_2153-2156, Xiao1987PRL_59_278-281, Polzik1992PRL_68_3020-3023,  Aasi2013NP_7_613-619,Pedrozo-Penafiel2020N_588_414-418,LIGOScientificCollaborationandVirgoCollaboration2021PRX_11_021053,Grote2013PRL_110_181101,Motes2015PRL_114_170802,Moller2017N_547_191-195,Colombo2022NP_18_925-930,Qin2023PRL_130_070801}. 
When $U$ is negative, the inclusion of the nonlinear Stark term significantly suppresses antibunching behavior, as shown in Figs.~\ref{fig1}(c)-(d). Specifically, at $U=-0.2$, the region where the system undergoes the double antibunching-to-bunching transition is significantly narrowed, as shown in Fig.~\ref{fig1}(c), compared to the $U>0$ case. As $U$ further decreases to $-0.5$,  shown in Fig.~\ref{fig1}(d), two key observations emerge: First, the primary antibunching effect weakens and shifts toward the DSC regime, with antibunching effect nearly vanishing. This signals a degradation of the double antibunching feature. Second, the photon bunching region expands significantly, approaching the DSC regime. 

In general, the nonlinear Stark coupling enhances two equally powerful but opposite effects: photon antibunching and photon bunching, both of which hold significant potential for quantum technologies. For positive Stark coupling $U > 0$, we observe a notable enhancement in both the richness and robustness of the antibunching effect, even in regimes where CRW terms dominate. In contrast, negative Stark coupling ($U < 0$) leads to significantly stronger bunching effects across a wide parameter range, providing a new degree of freedom to enhance signal contrast against noise.

\paragraph{Stark-coupled modulation of photon antibunching}

\begin{figure}[!htbp]
    \centering
\includegraphics[width=8.6cm]{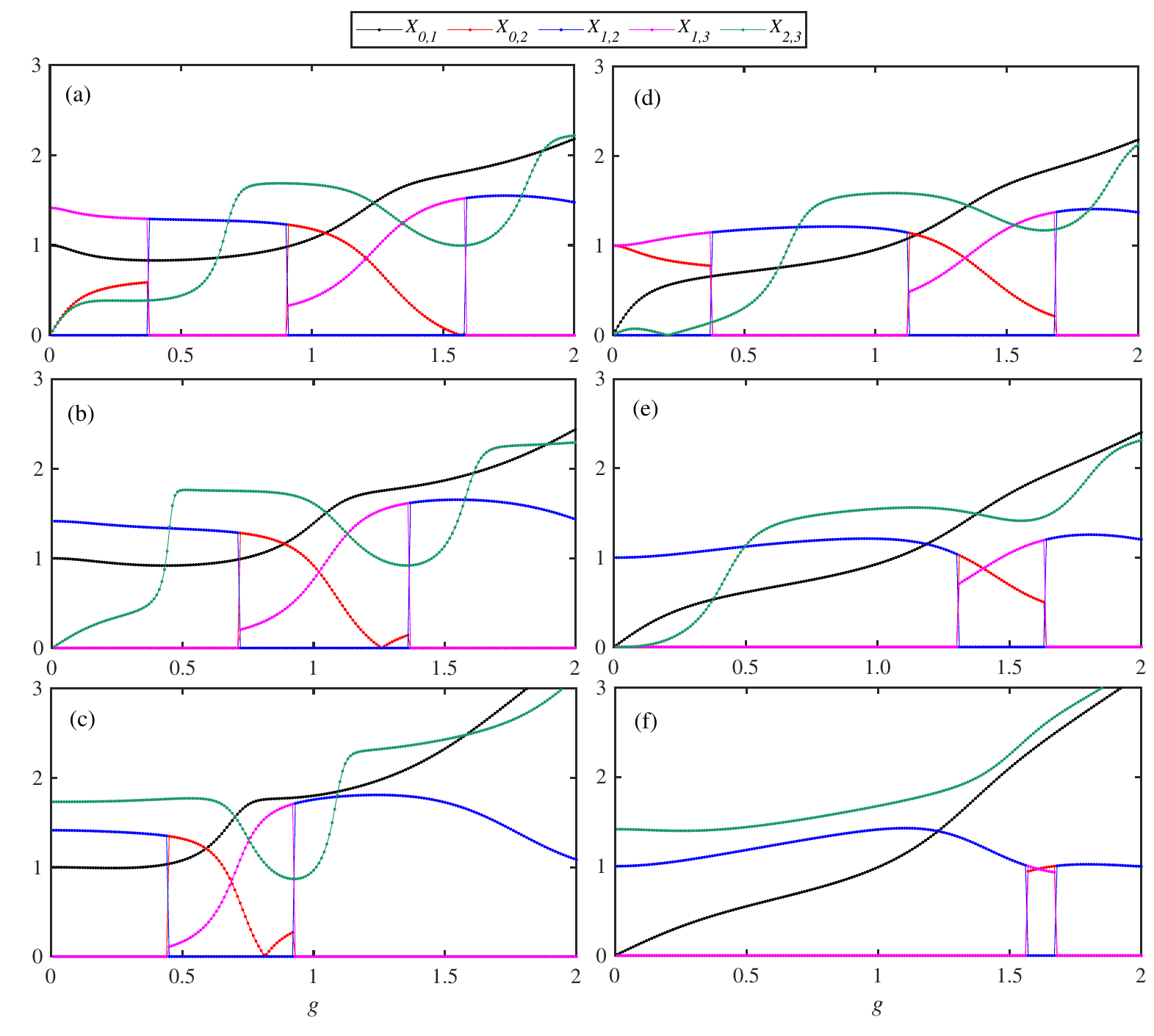}
    \vspace{-0.3cm}
    \caption{The matrix elements of the detection operator, $X_{ij}$, as a function of qubit-photon coupling strength $g$ for different Stark couplings: (a) $U = 0.2$, (b) $U = 0.5$, (c) $U = 0.8$, (d) $U = -0.2$, (e) $U = -0.5$, and (f) $U = -0.8$, with $r=0.2$. The Other parameters not mentioned here are the same as those in Fig.~\ref{fig1}.}
    \label{fig3}
\end{figure}

To understand the mechanism of antibunching effects controlled by the nonlinear Stark term, we analyze the evolution of matrix elements $X_{ij}$ for the detection operator in the dressed state, as shown in Fig.~\ref{fig3}. The pattern reveals frequent variations in photon transition pathways at $r=0.2$, with each switching event precisely coinciding with its first-order QPTs. Based on the quantum states involved in the first-order QPTs, we define two critical parameters: $g_c$, which marks the ground-state QPTs, and $\lambda_{c,n =1, 2}$, which characterizes transitions between excited states. According to  Xie~\textit{et~al.} \cite{xieFirstorderContinuousQuantum2020}, the critical coupling strengths between the ground state and the first excited state can be accurately identified as%
\begin{equation}
g_{c}=\sqrt{\frac{\Delta (1-U^{2})}{U(1+r^{2})+1-r^{2}}}.
\label{g_c(0)}
\end{equation}%
The critical coupling strength $\lambda_{c, n = 1,2}$ can be numerically determined from the recursive relations derived from the pole structure of the \textit{G} function. This correspondence indicates  a close relationship between the nonlinear Stark term's modulation  of the second-order correlation function and the system's QPTs signature. Specifically, for $U = 0.2$ in Fig.~\ref{fig3}(a), the corresponding phase transition points are $\lambda_{c,1}\approx 0.38$, $g_{c}\approx 0.91$, and $\lambda_{c, 2}\approx 1.59$, forming the essential foundation for subsequent analysis.

To clarify the mechanism, we present the approximate relation for the second-order correlation function given in Eq. \eqref{G2(0)}  
\begin{eqnarray}
\begin{aligned}
G_{2}(0) \approx &\{ (\Delta _{2,0}^{2}|X_{0,2}|^{2}+\Delta _{2,1}^{2}|X_{1,2}|^{2})\Delta
_{3,2}^{2}|X_{2,3}|^{2}P_{3}\\
&+\Delta _{1,0}^{2}|X_{0,1}|^{2}\Delta
_{3,1}^{2}|X_{1,3}|^{2}P_{3}\\
&+\Delta_{1,0}^{2}|X_{0,1}|^{2}\Delta _{2,1}^{2}|X_{1,2}|^{2}P_{2}\}/{\Delta _{1,0}^{4}|X_{0,1}|^{4}P_{1}^{2}}, 
\label{G2(0)A}
\end{aligned}
\end{eqnarray}%
where possibilities $P_n$ satisfy $P_0\gg P_1 \gg P_2\gg ...\gg P_n$. Before detailed analysis, we define $\eta _{i=1,2}$, where $\eta
_{1}=\Delta _{10}-\Delta _{21}$ and $\eta _{2}=\Delta _{10}-\Delta _{31}$, and the key features of photon statistics at distinct coupling regimes can be given as:
(i) For $g \leq \ \lambda_{c,1}$, $G_{2}(0)\approx 
\frac{\Delta _{3,1}^{2}|X_{1,3}|^{2}}{\Delta _{1,0}^{2}|X_{0,1}|^{2}}\mathrm{exp}(\eta _{2}/k_{\mathrm{B}}T)$; 

(ii) For $\lambda_{c,1} < g \leq g_{c}$, $G_{2}(0)\approx \frac{\Delta _{2,1}^{2}|X_{1,2}|^{2}}{\Delta_{1,0}^{2}|X_{0,1}|^{2}}\mathrm{exp}(\eta _{1}/k_{\mathrm{B}}T)$;

(iii) For $g_{c} < g < \lambda_{c,2}$, $G_{2}(0)\approx[{%
\Delta _{2,0}^{2}|X_{0,2}|^{2}\Delta _{3,2}^{2}|X_{2,3}|^{2}}\\
+{\Delta_{1,0}^{2}|X_{0,1}|^{2}\Delta_{3,1}^{2}|X_{1,3}|^{2}}]{\mathrm{exp}(\eta_{2}/k_{\mathrm{B}}T)}/{\Delta
_{1,0}^{4}|X_{0,1}|^{4}}$; 

(iv) For $g \geq \ \lambda_{c,2}$, $G_{2}(0)\approx\frac{\Delta _{2,1}^{2}|X_{1,2}|^{2}}{\Delta _{1,0}^{2}|X_{0,1}|^{2}}\mathrm{exp}(\eta _{1}/k_{\mathrm{B}}T)$.
We observe that processes (i) and (iii) involve two transition paths: $\left\vert \varphi _{3}\right\rangle \rightarrow \left\vert \varphi_{2(1)}\right\rangle \rightarrow \left\vert \varphi _{0}\right\rangle $, while processes (ii) and (iv) involve a single transition path: $\left\vert \varphi _{3}\right\rangle \rightarrow \left\vert \varphi_{2}\right\rangle \rightarrow \left\vert \varphi_{1}\right\rangle \rightarrow \left\vert \varphi _{0}\right\rangle $. A key metric for modulating photon statistics is the critical point at which antibunching effects first emerge, a threshold determined by parameter values $\eta _{i=1,2}$. When $|\eta _{i=1,2}|\gg k_{\mathrm{B}}T$, the system exhibits strong antibunching effects. Upon closer inspection, the positive Stark term enables antibunching to occur at smaller coupling strengths, persisting until $g_{f}\approx g_{c}$. Although $\lambda_{c,1}$ also represents a first-order transition point, it primarily affects the transition channels between states, due to the special form of Eq.~(\ref{G2(0)A}). Another key metric is the proportional distribution of coupling strength intervals corresponding to systems with single or multiple transition pathways. To better understand this aspect, we further examine the results under regimes where $U>0.2$. For instance, for $U=0.5$ in Fig.~\ref{fig3}(b) and $U=0.8$ in Fig.~\ref{fig3}(c),  process (i) disappears entirely compared to case of $U=0.2$, meaning that $\lambda_{c,1}$ vanishes. Generally, regions with multiple transition pathways, such as $\left\vert \varphi _{3}\right\rangle \rightarrow \left\vert \varphi
_{2(1)}\right\rangle \rightarrow \left\vert \varphi _{0}\right\rangle $, become inefficient, and the system predominantly exhibits a single-pathway two-photon process, i.e., $\left\vert \varphi _{3}\right\rangle \rightarrow \left\vert \varphi_{2}\right\rangle \rightarrow \left\vert \varphi_{1}\right\rangle \rightarrow \left\vert \varphi _{0}\right\rangle $. This indicates that the Stark coupling modulates  the  transition channels,  effectively controlling the extent of the photon-antibunching region.

\begin{figure}[!htbp]
    \centering
\includegraphics[width=8.6cm]{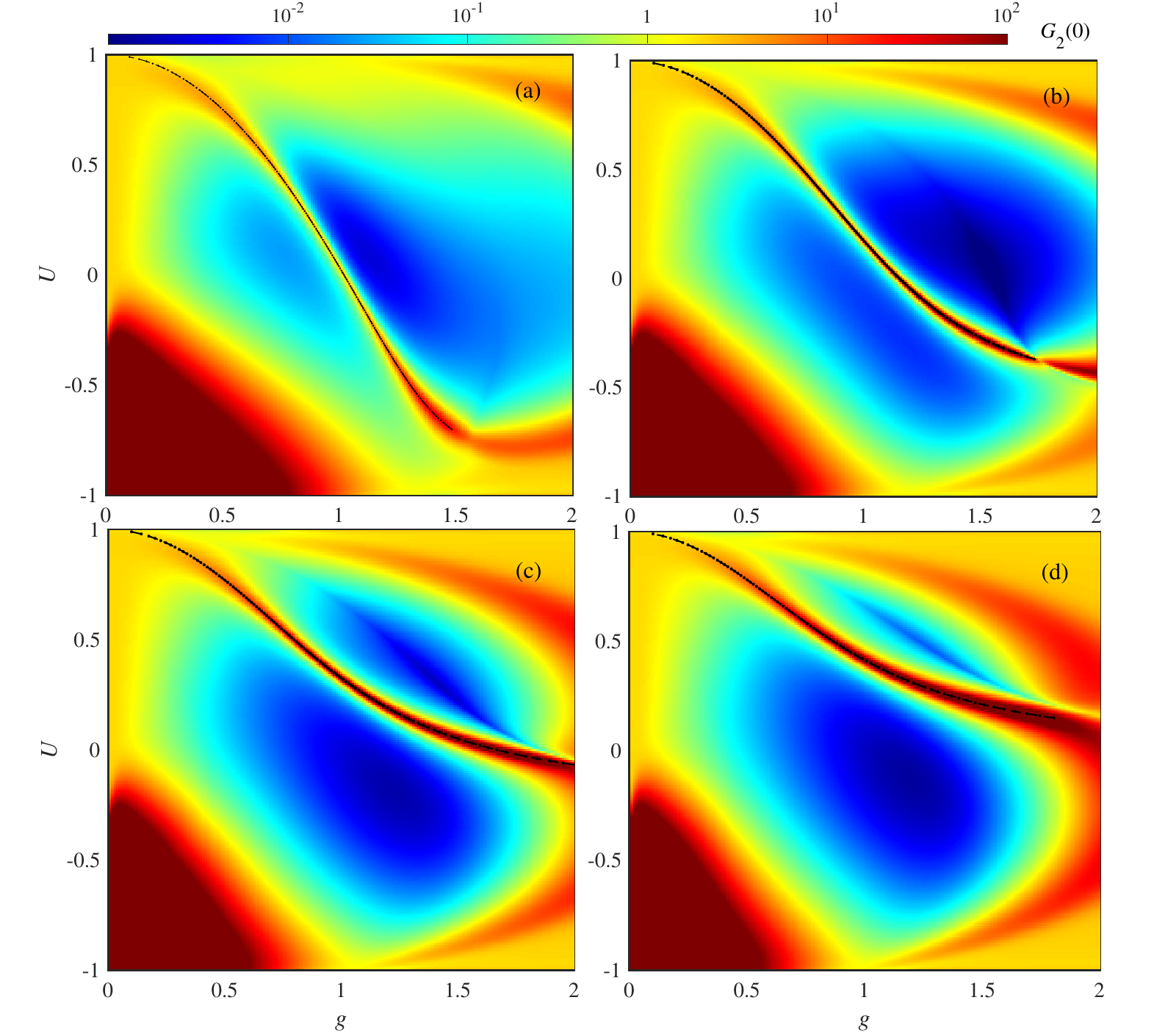}
    \vspace{-0.2cm}
    \caption{The zero-time delay two-photon correlation function, $G_2(0)$, as a function of the qubit-photon coupling strength $g$ and Stark coupling strength $U$ for different anisotropic parameters: (a) $r = 0.2$, (b) $r = 0.5$, (c) $r = 0.8$, and (d) $r = 1.0$. 
    The dotted-dashed lines indicate first-order QPTs in the ground state. Other parameters  not explicitly mentioned here are the same as those in Fig.~\ref{fig1}.}
    \label{fig2}
\end{figure}

Additionally, for the $U<0$ case, a new photon-bunching region emerges. For example, in Fig.~\ref{fig3}(d) with $U=-0.2$, we observe that $X_{0,1}$ increases slowly as the coupling strength $g$, where as it rises abruptly once $g$ becomes nonzero, as shown in Figs.~\ref{fig3}(a)-(c). Consequently, the second-order correlation function shows photon bunching across a wide range of coupling strengths, a result of the specific form of Eq.~(\ref{G2(0)A}). Similar behavior is observed in the Figs.~\ref{fig3}(e)-(f) as well. As $U$ decreases further, $X_{0,1}$ remains almost unchanged with increasing coupling strength, while the region exhibiting strong photon bunching expands, consistent with the results shown in Fig.~\ref{fig2}. Similar to the $U>0$ cases, we find that a larger $\left\vert U\right\vert $ leads to a greater deviation from AQRM behavior, enhancing single-channel processes and reducing multichannel ones, thereby shrinking the antibunching region. The antibunching region nearly disappears when $U$ decreases to -0.8, as shown in Fig.~\ref{fig2}.

The introduction of the nonlinear Stark term significantly alters the parity-controlled transition pathways, leading to the expanded single-path region. As a result, the parameter space exhibiting photon antibunching statistics is significantly contracted, which may have important implications for the experimental realization of single-photon~\cite{Kimble2008N_453_1023-1030,Scarani2009RMP_81_1301-1350} and multiphoton sources \cite{Munoz2014NP_8_550-555,Gou2024NJP_26_073046}.

\paragraph{Probing phase transitions with two-photon correlations}

  Level crossings are typically accompanied by parity symmetry breaking in the parity-conserving AQRSM, a hallmark of first-order QPTs. This corresponds to a discontinuity in the first derivative of energy with respect to the coupling strength $g$. As demonstrated in previous work of Xie~\textit{et al.} \cite{xieFirstorderContinuousQuantum2020}, the first-order phase transition point in the ground stat of AQRSM follows the explicit relationship in Eq.~\eqref{g_c(0)}, with results shown as dotted-dashed lines Fig.~\ref{fig2}. 

To refine the analysis, we focus on investigating the effect of the Stark term $U$ on the second-order correlation function for a given anisotropic parameter $r$, as shown in Fig.~\ref{fig2},  revealing its QPT correlation characteristics. By computing the second-order correlation function $G_{2}(0)$ using the quantum DME in Eq.~\eqref{DME}, we observe pronounced photon bunching and antibunching for varying anisotropic parameters $r=0.2, 0.5, 0.8, 1.0$ in Fig.~\ref{fig2}. Notably, the isotropic case in Fig. \ref{fig2}(d) provides the clearest evidence that the Stark coupling effectively controls photonic nonclassicality. Variations in the coupling strength $g$ continuously diversify photon statistics. According to Eq.~\eqref{G2(0)A}, the closure of $\Delta _{1,0}$ (i.e., the vanishing energy gap between the ground state and the first excited state)  triggers the first-order QPT,  leading to a nearly divergent red peak in $G_{2}(0)$. Moreover,  Fig.~\ref{fig2} shows multiple red peaks corresponding to photon bunching, which can be classified into two distinct types: (i) a photon-bunching region bounded by antibunching transitions, and (ii) bunching accompanied by antibunching on one side. 

Crucially, the first-type red bunching peak is fully bounded by antibunching features, forming a complete successive transition structure from “antibunching to bunching to antibunching”. This characteristic peak shows perfectly matches the first-order phase transition points $g_{c}$ from Eq.~\eqref{g_c(0)}, offering a clear experimental protocol for QPT detection. In contrast, the second-type bunching peak appears only in the negative Stark parameter regime and is not related to the phase-transition behavior. These results establish the successive transition signature of $G_2(0)$ as a definitive diagnostic marker for first-order QPTs in quantum light-matter interacting systems.

\subsection{Higher-order photon correlations with a tunable Stark field}

The previous subsections have established the successive transition signature of the second-order correlation function as a probe for experimentally detecting QPTs.	In this subsection, we use the higher-order correlation function to systematically investigate two key aspects: (1) the identification of phase transitions that are undetectable by second-order correlations within the parameter space, and (2) the exploration of mechanisms for multiphoton correlations controlled by the tunable nonlinear Stark coupling.

\begin{figure}[!htbp]
    \centering
\includegraphics[width=8.6cm]{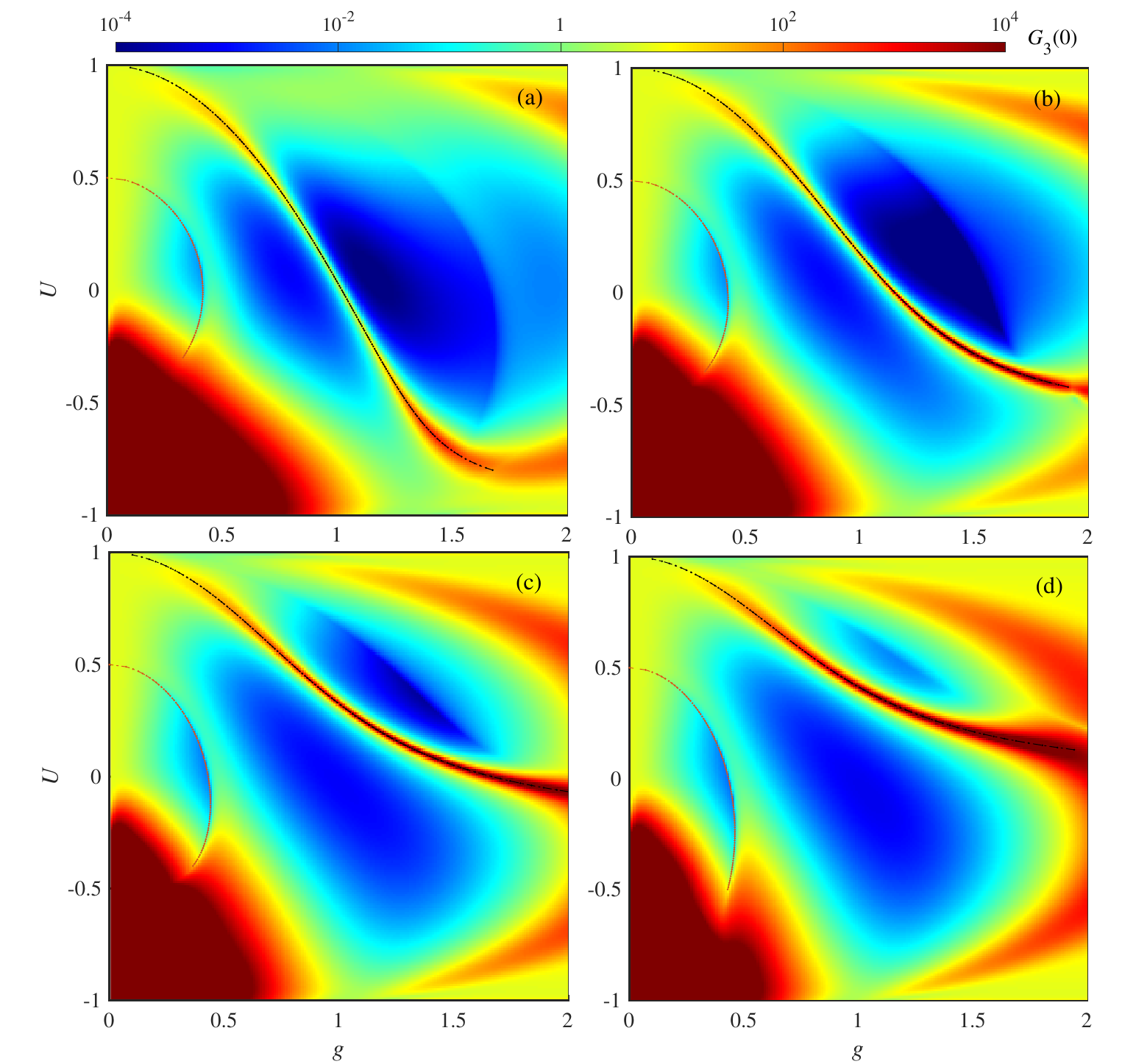}
    \vspace{-0.2cm}
    \caption{The zero-time delay three-photon correlation function, $G_{3}(0)$, as a function of the Stark coupling strength $U$ and the coupling strength $g$ for different anisotropic parameters: (a) $r = 0.2$, (b) $r = 0.5$, (c) $r = 0.8$, and (d) $r = 1.0$. Black dotted-dashed lines indicate ground-state first-order QPTs, while dark orange dotted-dashed lines represent the level crossing between the second and third excited states. All other parameters explicitly mentioned are the same as those in Fig.~\ref{fig1}.}
    \label{fig4}
\end{figure}

\paragraph{Three-photon correlation signature}
 The primary contribution to $G_{3}(0)$, as derived from Eq. \eqref{G3(0)},
is the multiphoton transition pathway $\left\vert \varphi
_{3}\right\rangle \rightarrow \left\vert \varphi _{2}\right\rangle
\rightarrow \left\vert \varphi _{1}\right\rangle \rightarrow \left\vert
\varphi _{0}\right\rangle $.
The approximate expression is given by
\begin{equation}
G_{3}(0)\approx \frac{\Delta_{2,1}^{2}\Delta_{3,2}^{2}\left\vert
X_{1,2}\right\vert ^{2}|X_{2,3}|^{2}}{\Delta_{1,0}^{4}|X_{0,1}|^{4}}\exp
(\eta _{3}/k_{\mathrm{B}}T),\label{G_3(0)A}
\end{equation}%
where $\eta _{3}=2\Delta _{10}-\Delta _{21}-\Delta _{32}$ represents the effective level separation. The vanishing energy gap between the ground state and the excited states, i.e. $\Delta_{1,0} = 0$, still leads to a near-divergent behavior in higher-order correlation functions. Furthermore, higher-order correlation functions depend on the energy gap $\Delta_{1,0}$ in a higher-order manner, leading to more pronounced statistical signatures compared to second-order correlation functions. 

We present the behavior of third-order correlation functions as a function of both the nonlinear Stark coefficient $U$ and the coupling strength $g$, for fixed anisotropic parameters $r=0.2,0.5,0.8,1.0$, as shown in Fig.~\ref{fig4}. We observe that, except in certain regions, the three-photon correlation function is stronger in areas where the two-photon correlation is pronounced, and weaker where the two-photon correlation is less apparent.

For any anisotropic parameter $r$, the third-order correlation also exhibits two types of bunching, in agreement with the second-order results. One of these, the photon-bunching region bounded by antibunching transitions, also accurately characterizes first-order QPTs in the ground state, perfectly aligning with the analytically derived black dotted-dashed lines in Fig.~\ref{fig4}. Another bunching phenomenon, which shows no phase transition (similar to that in Fig.~\ref{fig2}), also occurs, suggesting an intrinsic connection between the distinct-order correlation functions. We also identify a new stripe-shaped region in Fig.~\ref{fig4}, where the third-order correlation function displays pronounced antibunching effects, accompanied by bunching effects on one side. This reveals another successive transition signature that extends beyond the ground-state QPTs depicted in Fig.~\ref{fig2}. This feature precisely coincides with the level crossing between the second and third excited states, as clearly shown by the dark orange dotted-dashed lines in Fig.~\ref{fig4}.

The third-order correlation function uncovers deeper statistical properties of the photons, characterized by more pronounced photon bunching and antibunching.	In addition to capturing first-order QPTs in the ground state, the third-order correlation function also reveals their excited-state counterparts, manifested through its distinct successive transition signature. Higher-order correlation functions will also naturally uncover additional hidden statistical properties.

\paragraph{Multiphoton transition processes}

We systematically analyze the influence of the Stark coupling on the third-order correlation function $G_{3}(0)$ and its underlying physical processes, as shown in Fig.~\ref{fig4}(a) for $r=0.2$. This formulation in Eq.~\eqref{G_3(0)A} highlights  the crucial role of parity selection rules in three-photon processes. Specifically, the transitions from the ground state $\left\vert \varphi
_{0}\right\rangle $ and the first excited state $\left\vert \varphi
_{1}\right\rangle $, and between $\left\vert \varphi _{2}\right\rangle $ and $\left\vert \varphi _{3}\right\rangle $, always follow the parity selection rules. The key factor governing the third-order photon correlation function is whether the parity between $ \left\vert \varphi _{1}\right\rangle $ and $\left\vert \varphi_{2}\right\rangle $ matches.

A detailed analysis of the altered transition pathways near the critical point in Fig.~\ref{fig3}(a) reveals that: 

(i) For $g\leq \lambda_{c,1}$, the identical parity between $\left\vert \varphi_{1}\right\rangle $ and $\left\vert \varphi _{2}\right\rangle $ causes $X_{1,2}=0$, thereby suppressing the three-photon processes, even though $\eta _{3}$ remains large; 

(ii) As the coupling strength increases to the range $\lambda_{c,1}<g<g_{c}$, the parity switching between $\left\vert \varphi _{2}\right\rangle $ and $\left\vert \varphi_{3}\right\rangle $ significantly enhances the initial three-photon transition probability;

(iii) As the system enters the regime $g_{c}\leq g< \lambda_{c,2}$, the parity transition between $\left\vert\varphi_{0}\right\rangle $ and $\left\vert \varphi _{1}\right\rangle $ leads to $\Delta_{1,0}=0$ in $g = g_{c}$, causing a divergent growth of $G_{3}(0)$. Upon crossing this ground-state phase transition point $g_{c}$, the third-order correlation function sharply drops to values significantly lower than those observed before the phase transition, due to the disappearance of $X_{1,2}$;

(iv) As the system enters the regime $g \geq \lambda_{c,2}$, the three-photon processes are further enhanced due to parity switching between states $\left\vert\varphi_{2}\right\rangle $ and $\left\vert \varphi _{3}\right\rangle $. 

The behavior of $X_{1,2}$ in Fig.~\ref{fig3} clearly illustrates how increasing $\left\vert U\right\vert $ progressively expands the parameter range for the transition between the first and second excited states, highlighting the essential role of the nonlinear Stark term in controlling three-order correlation functions. Similar behavior is also observed in higher-order correlation functions, where transitions between specific states are crutical, as they directly determine the scale of multiphoton processes. These transitions are closely related to parity. The Stark coupling precisely modulates the magnitude of the corresponding correlation functions by adjusting the regions in which these transitions occur. The nonlinear Stark coupling offers a new avenue for controlling multiphoton sources and $n$-photon blockade through higher-order correlation features.

\section{Quadrature squeezing of photons}\label{sec:squeezing}
In this section, we focus on the well-known phenomenon of quadrature squeezing, which occurs when the quantum fluctuation of one quadrature component of an electromagnetic field falls below the vacuum limit.	Simultaneously, the fluctuation of the conjugate quadrature component increases to comply with the Heisenberg uncertainty principle. Quadrature squeezing is of significant importance in precision measurement~\cite{Caves1980RMP_52_341-392b, Grangier1987PRL_59_2153-2156, Xiao1987PRL_59_278-281, Polzik1992PRL_68_3020-3023,  Aasi2013NP_7_613-619,Pedrozo-Penafiel2020N_588_414-418,LIGOScientificCollaborationandVirgoCollaboration2021PRX_11_021053,Grote2013PRL_110_181101}, quantum communication \cite{Cleve1997PRA_56_1201-1204, Braunstein2005RMP_77_513-577, Kimble2008N_453_1023-1030, Scarani2009RMP_81_1301-1350, Wang2019PRA_99_042309}, and quantum metrology \cite{Motes2015PRL_114_170802,Moller2017N_547_191-195,Colombo2022NP_18_925-930,Qin2023PRL_130_070801}. This section analyzes the effects of the nonlinear Stark coupling $U$ on quadrature squeezing, offering theoretical insights for the design and optimization of quantum optical devices.

\subsection{Quadrature squeezing}

According to Heisenberg's uncertainty principle $(\Delta A)(\Delta B)\geq \frac{1}{2}\left\vert \left\langle [A,B]\right\rangle
\right\vert $ with $\Delta \mathcal{O}=\sqrt{\left\langle \mathcal{O}^{2}\right\rangle
-\left\langle \mathcal{O}\right\rangle ^{2}}$, we conclude that two physical quantities can be simultaneously measured only if they commute completely, meaning they share a common eigenstate. Otherwise, precise simultaneous measurement of both quantities is fundamentally impossible. For operators of the quadrature component of the electromagnetic field, defined as $X=a+a^{\dag },P=\frac{1}{i}(a-a^{\dag })$, the vacuum state $\left\vert0\right\rangle $ is a minimum uncertainty state. These  intrinsic quantum fluctuations, known as the quantum noise limit, satisfy $\Delta X=1, \Delta P=1$. Coherent states $\left\vert \alpha \right\rangle $ also saturate this quantum noise limit. In contrast, the squeezed states $\left\vert \xi \right\rangle =S(\xi )\left\vert
0\right\rangle $, generated by the squeezing operator $S(\xi )=\exp [\frac{1}{2}(\xi ^{\ast }a^{2}-\xi a^{\dag 2})]$ with $\xi =r\exp (2i\theta )$, introduce a phase-dependent rotation of the quadrature axes. This leads to the redefinition of the quadrature components as 
\begin{equation}
X_{\theta }=ae^{-i\theta }+a^{\dag }e^{i\theta },P_{\theta }=\frac{1}{i}%
(ae^{-i\theta }-a^{\dag }e^{i\theta }),  \label{XP}
\end{equation}%
yielding%
\begin{equation}
\Delta X_{\theta }=e^{-r},\Delta P_{\theta }=e^{r}. 
\end{equation}
Thus, squeezed states remain minimum uncertainty states but they redistribute noise unevenly between the quadratures ~\cite{Stoler1970PRD_1_3217-3219,Stoler1971PRD_4_1925-1926}. Here, $\theta =0$ corresponds to $%
X_{\theta }=X$, while $\theta =\frac{\pi }{2}$ gives $X_{\theta }=P$. The degree of principal quadrature squeezing is quantified by the parameter \cite{Ma2011PR_509_89-165}
\begin{equation}
\xi _{\mathrm{B}}^{2}=\min_{\theta \in [0,2\pi )}(\Delta X_{\theta })^{2}.  \label{QS}
\end{equation}
The relation $\xi_{\mathrm{B}}^{2}=1$ holds  for both the vacuum and the coherent states, reflecting their traditional uncertainty bounds. Photon squeezing occurs when the condition $\xi_{\mathrm{B}}^{2}<1$ is met, whereas no squeezing is observed when $\xi_{\mathrm{B}}^{2} {\ge} 1$.
\subsection{Photon squeezing modulated by nonlinear Stark coupling}

\begin{figure}[!htbp]
    \centering
\includegraphics[width=8.6cm]{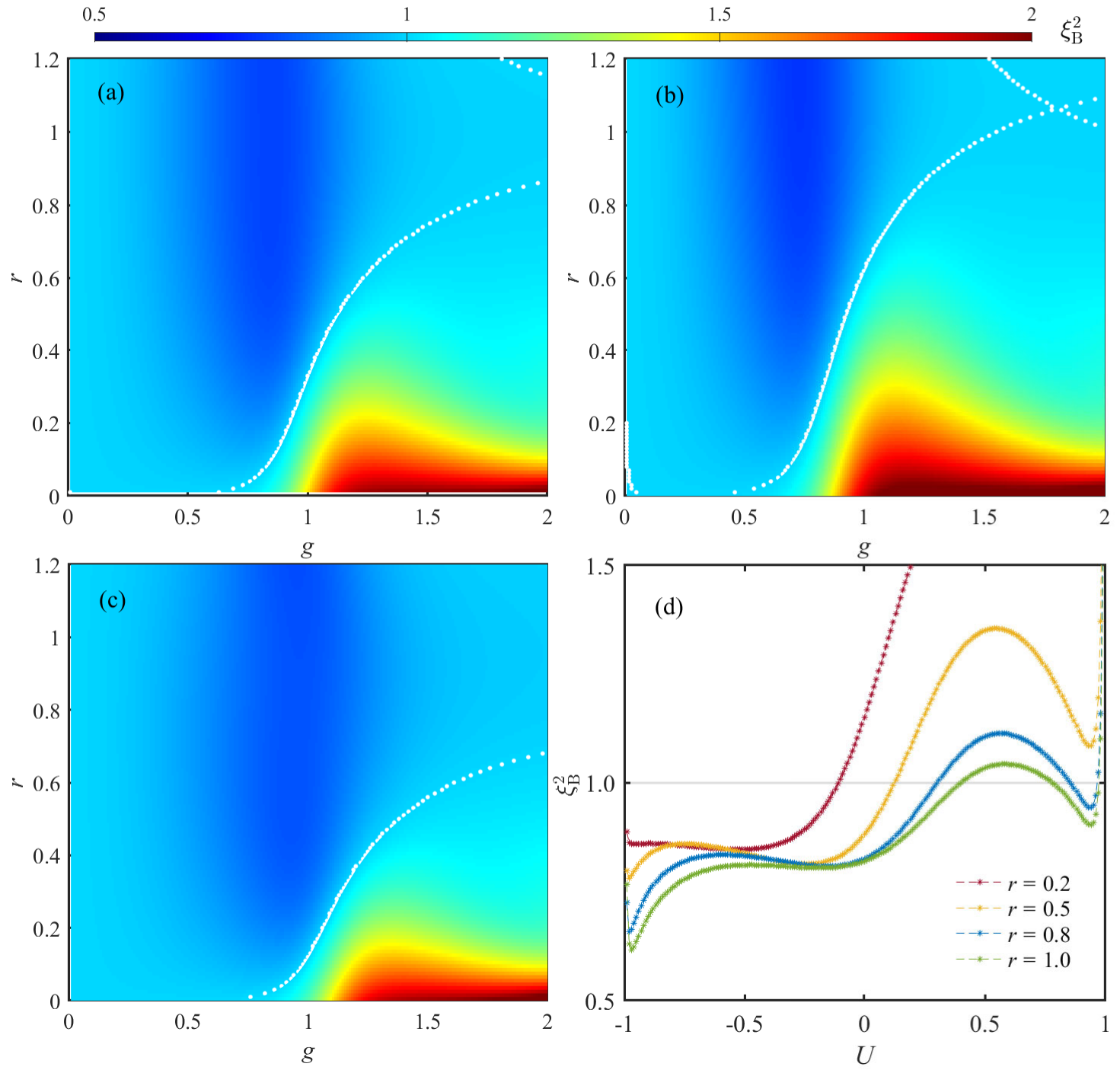}
    \vspace{-0.3cm}
    \caption{The quadrature squeezing factor, $\xi _{\mathrm{B}}^{2}$, as a function of anisotropic parameters $r$ and the coupling strength $g$ for different Stark coupling strengths: (a) $U = 0$, (b) $U = 0.2$, and (c) $U = -0.2$. The region \textcolor{blue}{almost} located to the left of the white dotted lines represents the area where $\xi_{\mathrm{B}}^{2}<1$, indicating photons squeezing. (d) The quadrature squeezing factor, $\xi _{\mathrm{B}}^{2}$, as a function of the Stark term $U$ at coupling strength $g = 1$ for various anisotropic parameters $r$. All other parameters not mentioned here are the same as those in Fig.~\ref{fig1}.}
    \label{fig5}
\end{figure}

Fig.~\ref{fig5}(a) illustrates the quadrature squeezing factor of the AQRM, specifically without the nonlinear Stark coupling. The region where $\xi_{\mathrm{B}}^{2} < 1$, located to the left of the white dotted lines, represents the parameter space in which the photons exhibit squeezing. Photon squeezing persists over a wide range of anisotropic parameters $r$. In the RW-dominated regime $r{\ll}1$, the squeezing region expands with increasing $r$. In contrast, the squeezing region contracts as $r$ increases in the CRW-dominated regime ($r > 1$). Although the qubit-photon coupling strength corresponding to the maximum photon squeezing varies slightly with the anisotropic parameter $r$, the maximum photon squeezing consistently stabilizes around $g\approx0.8$. Additionally, the parameter region exhibiting photon squeezing has notable limitations: (i) In the DSC regime ($g>1$,) most parameter sets do not exhibit apparent photon squeezing; (ii) In the USC regime ($0.1\leq g<1$), a parameter range without photon squeezing emerges when $r<0.3$.
Next, we introduce the nonlinear Stark coupling to enhance control over photon quadrature squeezing, as shown in Figs. \ref{fig5}(b) for $U=0.2$ and (c) $U=-0.2$, which display the variation patterns of the quadrature squeezing factor. For the $U>0$, the parameter range for squeezing shrinks, whereas for the $U<0$, new squeezing emerges in the USC and DSC regions where AQRM originally exhibited no squeezing. To quantitatively characterize the influence of the Stark term, we examine the quadrature squeezing factor in Fig.~\ref{fig5}(d) for a fixed coupling strength $g=1$, varying the anisotropic parameter $r$. The result reveals a systematic pattern: photon squeezing consistently emerges when the Stark term $U$ is negative, with this quantum effect persisting even in the small positive-Stark regime. Specifically, for larger anisotropic parameters ($r=0.8$) and (1.0), photon squeezing vanishes with increasing $U$, and revives at stronger Stark coupling.
%This demonstrates the jointly affect of the Stark term and the anisotropic parameter on producing rich squeezing behaviors, which we further explore in the \hyperref [app:supp] {Appendix}. 

In conclusion, we have elucidated the influence of Stark coupling $U$ on quadrature squeezing in the AQRSM, highlighting its role in introducing a greater diversity of photonic phenomena. These effects primarily manifest through: (i) a modification of the accessible parameter space for photon squeezing, and (ii) precise control of the squeezing strength at specific parameter configurations. 

\subsection{Intuitive expression of quadrature squeezing}
To gain a deeper understanding of how the nonliner Stark term $U$ and the anisotropy term $r$ influence  photon squeezing, we first simplify the quadrature squeezing factor in Eq.~\eqref{QS} as%
\begin{equation}
\xi _{\mathrm{B}}^{2}=2[\left\langle a^{\dag }a\right\rangle -\left\langle a^{\dag
}\right\rangle \left\langle a\right\rangle -2\frac{\mathrm{Var}(a)\mathrm{Var}(a^{\dag })}{%
\mathrm{Var}(a)+\mathrm{Var}(a^{\dag })}]+1,
\end{equation}%
where $\mathrm{Var}(\mathcal{O})=\left\langle \mathcal{O}^{2}\right\rangle -\left\langle \mathcal{O}\right\rangle ^{2}$. Based on the symmetry properties of the system $\left\langle a^{\dag }\right\rangle
=\left\langle a\right\rangle =0$ and $\mathrm{Var}(a)=\mathrm{Var}(a^{\dag })$, the expression can be further simplified to%
\begin{equation}
\xi _{\mathrm{B}}^{2}=2[\left\langle a^{\dag }a\right\rangle -\left\langle
a^{2}\right\rangle ]+1.  \label{xi_new}
\end{equation}%
This simplified form provides profound physical insight: the emergence and disappearance of the photon squeezing are fundamentally determined by the relative magnitudes of $\left\langle a^{\dag }a\right\rangle $ and $%
\left\langle a^{2}\right\rangle (\left\langle a^{\dag 2}\right\rangle )$. In other words, photon squeezing occurs with a strong two-photon process but is absent when the process is weak.
\begin{figure}[!htbp]
    \centering
\includegraphics[width=8.6cm]{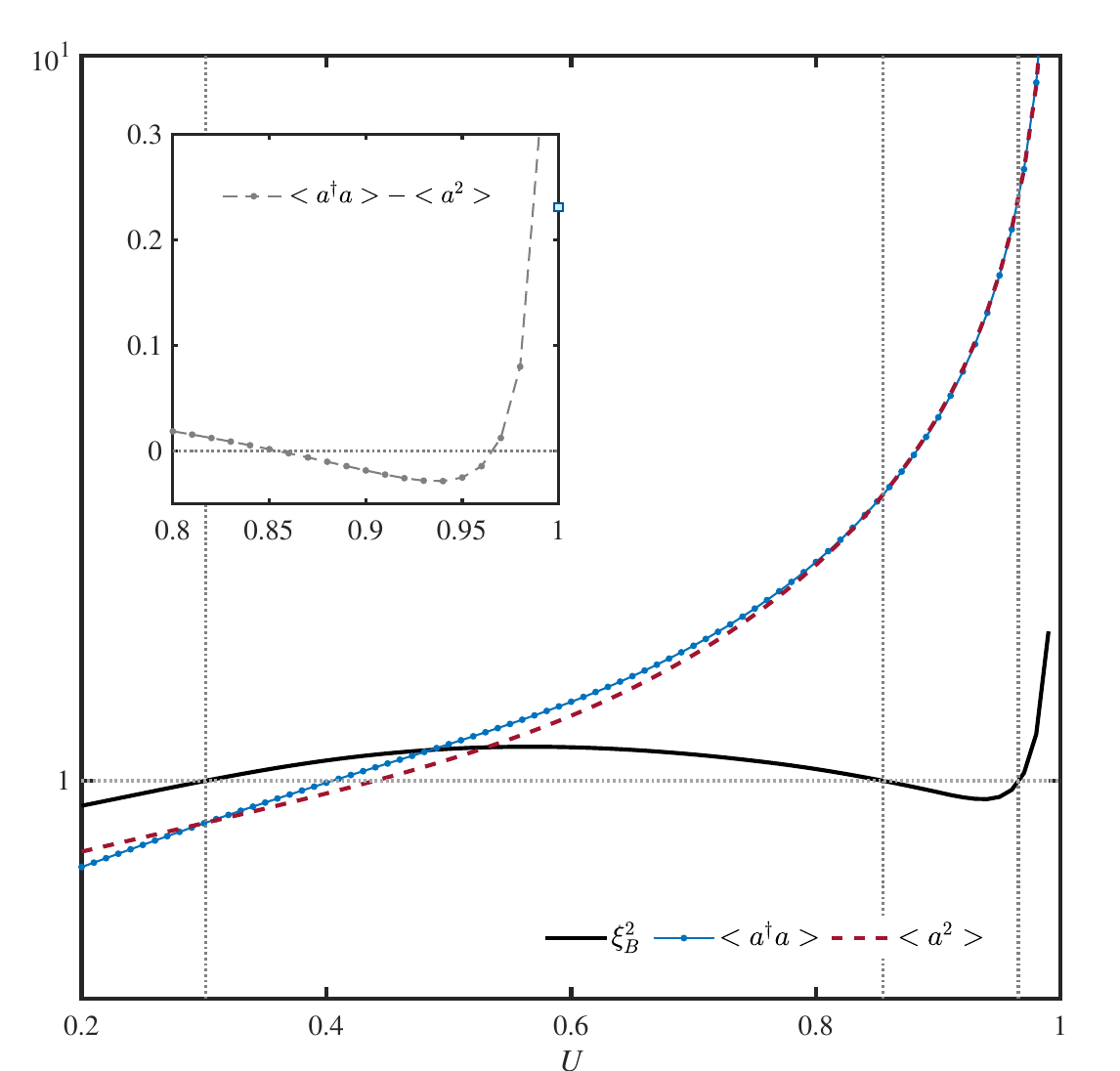}
    \vspace{-0.5cm}
    \caption{The quadrature squeezing factor $\xi _{\mathrm{B}}^{2}$, $\left\langle a^{\dag }a\right\rangle $, and $\left\langle a^{2}\right\rangle$ in AQRSM as a function of Stark coupling strength $U$ at coupling strength $g=1.0$ and anisotropic parameter $r=0.8$, with the inset showing  the difference between $\left\langle a^{\dag }a\right\rangle $ and $\left\langle a^{2}\right\rangle$ in the range $U=0.8$ to 1.0. Other parameters are the same as those in Fig.~\ref{fig5}.}
    \label{fig6}
\end{figure}

For the case $r=0.8$ in Fig.~\ref{fig5}(d), we present the relationship between $\xi_\mathrm{B}^2$, $\left\langle a^{\dag }a\right\rangle $, and $\left\langle a^{2}\right\rangle$ in Fig. \ref{fig6}, which reveals that the relative variations between $\left\langle a^{\dag }a\right\rangle $ and $\left\langle a^{2}\right\rangle$ fundamentally govern the alternating appearance of squeezed and non-squeezed photon states. As the Stark coupling increases, the photons exhibit several transitions between squeezing and no squeezing. The transition points correspond precisely to the intersections of the curves representing $\left\langle a^{\dag }a\right\rangle $ and $\left\langle a^{2}\right\rangle$. This discovery not only establishes a quantitative relationship between the Stark coupling and the photon squeezing, but, more importantly, reveals the physical mechanism for precise control of quadrature squeezing by regulating the dynamic balance between the photon number fluctuations and the two-photon process, offering new theoretical insight into quadrature squeezing.

\section{Conclusion and outlook}\label{sec:Conclusion}
In this work, we systematically investigate the nonclassical photonic properties and quadrature squeezing in AQRSM using the quantum DME approach, highlighting the crucial role of the Stark coupling in mediating these quantum phenomena. The principal findings are summarized as follows:

First, we perform a comprehensive investigation of photon antibunching and its control mechanisms in the AQRSM, using both second- and higher-order correlation functions. The inclusion of Stark coupling substantially enriches both photon antibunching and bunching phenomena, with the effects critically dependent on the sign and magnitude of the Stark coupling. By precisely manipulating photon transition pathways enabled by Stark coupling, we achieve effective control over the emergence of photon antibunching.	This approach establishes a novel protocol for implementing photon blockade and facilitating the preparation of entangled states.

Second, we confirmed that first-order QPTs and level crossings between excited states in the AQRSM can be accurately captured through the successive transition signatures of second- and third-order correlation functions.	Analysis of correlation functions reveals distinct signatures of QPTs: (i) A photon-bunching region bounded by antibunching transitions indicates a successive transition behavior, characteristic of first-order QPTs in the ground state; (ii) Conversely, strongly pronounced antibunching accompanied by bunching phenomena on one side provides clear evidence of the level crossing between excited states. These findings lay the foundation for predicting quantum criticality in novel systems. Furthermore, considering higher-order correlation functions is expected to reveal more intricate properties of the system.

Finally, this study demonstrates that nonlinear Stark coupling enables effective control of quadrature squeezing, governing both its accessible parameter space and ultimate achievable strength. As confirmed in this work, the emergence of these effects arises from the system's ability to establish a dynamical equilibrium between the photon population and two-photon processes. The proposed mechanism demonstrates significant potential for applications in quantum-enhanced metrology and advanced communication protocols.

In short, these results demonstrate that Stark coupling acts as a pivotal control parameter, allowing the photon to display a rich variety of nonclassical properties and quadrature squeezing in the dissipative AQRSM system. This opens new avenues for quantum information processing and advanced quantum technologies within strongly coupled light–matter systems.

\begin{acknowledgments}
 This work is supported by the National Key R\&D Program of China under Grant No. 2024YFA1408900 (YXZ and QHC) and the Zhejiang Provincial Natural Science Foundation of China under Grant No. LZ25A050001 (CW).
\end{acknowledgments}


\bibliography{apssamp}

\end{document}